\documentclass[review]{elsarticle}

\usepackage{graphicx}
\usepackage{color}
\usepackage{lineno,hyperref}

\usepackage{amssymb}

\journal{Nuclear Physics A}

\begin{document}

\begin{frontmatter}



\title{Impact of the chromo-electromagnetic field fluctuations on transport coefficients of heavy quarks and shear viscosity to entropy density ratio of quark-gluon plasma}


\author{\it Ashik Ikbal Sheikh}
\ead{ashikhep@gmail.com}
\author{\it Zubayer Ahammed}
\ead{za@vecc.gov.in}

\address{Variable Energy Cyclotron Centre, HBNI, 1/AF Bidhan Nagar, Kolkata 700 064, India}

\begin{abstract}
The chromo-electromagnetic field fluctuations in the quark-gluon plasma (QGP) play an important role as these field fluctuations result  energy gain of heavy quarks. We consider these fluctuations and evaluate the transport coefficients, e.g., drag and diffusion coefficients of charm quarks and shear viscosity to entropy density ratio ($\eta/s$) of the QGP. We find a significant effect of such fluctuations on the transport coefficients. These fluctuations cause a reduction of the drag and diffusion coefficients. We also observe that the shear viscosity to entropy density ratio of the QGP is closer to the value obtained in Lattice QCD (LQCD) and functional renormalization group calculations when the effects of such fluctuations are included. 

\end{abstract}

\begin{keyword}
\texttt{Quark-Gluon Plasma, Chromo-electromagnetic Field Fluctuations, Heavy Quarks, Shear viscosity to entropy density ratio}

\end{keyword}

\end{frontmatter}


\section{Introduction}
\label{intro}

 One of the main emphasis of present day heavy-ion experiments is to characterize quark-gluon plasma (QGP) or more precisely to determine its transport coefficents\cite{Muller13}. Immediately after the creation, the QGP will be cooled by expansion due to large internal pressure and will revert to the hadronic phase. The existence of these two phases (the deconfined QGP phase and hadronic phase) has been confirmed by recent Lattice QCD (LQCD) calculations\cite{Borsanyi12,Bazavov,Borsanyi14} and experimental observations\cite{br,ph,phn,str,al}. 
 During the transition from the deconfined QGP phase to the hadronic phase, the system may encounter critical point in QCD phase diagram. The characterization of the medium at critical point is one of the most 
 challenging problems in heavy-ion collisions at the relativistic energies. The LQCD calculations indicate 
 that the transition occurs near the critical temperature around $T_{c} \sim 155 $ MeV \cite{Karsch} at zero baryonic chemical potential.
 
 Amongst many, one of the efficient probe to understand the QGP is the heavy quarks which are mostly produced from the fusion of partons at the early stage of heavy-ion collisions.  Due to their large mass, few heavy quarks are produced at the later stage and none in the hadronic matter. This helps them to play a crucial role to characterize this deconfined medium formed in such collisions. The heavy quarks, after their production, propagate through the medium and lose energy throughout their path of propagation. The energy loss suffered by the heavy quarks is reflected in the relative suppression of heavy-flavoured hadrons\cite{ALICE_D,CMS_D_2TeV,CMS_D_5TeV,CMS_B_5TeV}. Heavy quarks lose energy by interacting with the light partons of the thermal background (QGP) and  by radiating gluons, {\it viz.}, bremsstrahlung process due to the deceleration of the heavy quarks. The collisional\cite{TG,BT,Alex,PP} and radiative\cite{MG,mgm05,dokshit01,dead,wicks07,GLV,armesto2,B.Z,W.C,Vitev,AJMS,KPV} energy loss have been reported extensively by several authors. 
 
 In general field fluctuations are not considered when energy loses are calculated in the QGP. Since the  QGP is a statistical system  of  coloured partons that are moving randomly, widespread stochastic fluctuations are  also expected in the system. These microscopic fluctuations  affect the response of the system on the influence of external perturbations.  The impact of such electromagnetic field fluctuations on the propagation of charged particles though a non-relativistic classical plasma has been estimated by several authors\cite{Gasirowicz,Sitenko,Akhiezer,Kalman,Thompson,Ichimaru}. 
 The effective parton energy loss while propagating in the QGP considering the effect of stimulated gluon emission and thermal absorption has been reported in Ref\cite{wangwang}. On the other hand the effect of chromo-electromagnetic field fluctuations in the QGP cause energy gain of heavy quarks of all momenta, significantly at lower momentum \cite{Fl}. This is due to the fact that the statistical change in the energy of the propagating heavy quarks takes place when the chromo-electromagnetic field fluctuations cause the fluctuations in the velocities of heavy quarks. This  energy gain results in reduction of the total energy loss. The effect of this energy gain is not negligible and it has an important impact on the suppression of heavy-flavoured hadrons which was shown in our previous work\cite{Ours}. 
 
 Generally, the magnitudes of transport coefficients are determined by the interaction of the heavy quarks in the medium. Therefore, the estimation of transport coefficients of the QGP  using heavy quarks is a field of immense interest. In earlier works\cite{BS,Rapp,MGM,Sur,Gos,Sur11,SKD10,Vitev7}, while calculating drag and diffusion coefficients of heavy quarks and shear viscosity to entropy density ratio ($\eta/s$) of the QGP, the energy gain due to field fluctuations was not considered, only collisional and radiative loss processes were considered. In this work, we consider the effect of chromo-electromagnetic field fluctuations and estimate the drag and diffusion coefficients of charm quarks and $\eta/s$ of the QGP medium. We find a significant effect of these field fluctuations.
 
 The present article is organized as follows: In the next section, we briefly outline the space-time evolution for the QGP medium and initial conditions, the model for heavy quarks energy loss and the effect of chromo-electromagnetic field fluctuations. We consider the collisional energy loss of heavy quarks by Brateen and Thoma (BT) formalism\cite{BT} and the radiative energy loss by reaction operator formalism (DGLV)\cite{wicks07,GLV} and the energy gain due to chromo-electromagnetic field fluctuations as prescribed in Ref.\cite{Fl}. In Sec.\ref{sec3}, we discuss the formalism for heavy quarks drag and diffusion in the QGP medium. The shear viscosity to entropy density ratio ($\eta/s$) of the QGP is estimated by using  diffusion coefficients of charm quarks and it has been discussed in Sec.\ref{sec4}. Sec.\ref{sec5} is devoted to summary and conclusion.

\section{Methodology}
\label{sec2}
\subsection{Initial condition and space-time evolution}
\label{IC}

We consider a heavy quark produced at transverse position $\vec{r}$ with an angle $\phi$ relative to the radial direction $\hat{r}$ in a heavy-ion 
collision, propagating through the QGP medium and losing energy. As far as the energy loss is concerned, the total path traversed by the heavy quark is an important quantity to be estimated. The path length $L$ traveled by the heavy quark inside the medium is calculated as\cite{Muller}:
\begin{equation}
L(r,\phi) = \sqrt{R^{2}-r^{2}\sin^{2}{\phi}} - r\cos{\phi}.
\end{equation}
where  $R$ is the radius of the colliding nuclei. The average distance traveled by the heavy quark inside the QGP is
\begin{eqnarray}
\label{eq2}
\langle L \rangle &= \frac{\int\limits_{0}^{R}rdr\int\limits_{0}^{2\pi}L(r,\phi)T_{AA}(r,b)d\phi}{\int\limits_{0}^{R}rdr\int\limits_{0}^{2\pi}T_{AA}(r,b)d\phi} , 
\end{eqnarray}
where $T_{AA}(r,b)$ is the nuclear overlap function at an impact parameter $b$, obtained from Glauber Model calculation as, $T_{AA}(r,b) = \rho(|\vec{r}|)\rho(|\vec{r} - \vec{b}|)$ (with $\rho(|\vec{r}|)$ is the density of nucleus assumed to be a sharp sphere with radius $R = 1.1 A^{1/3}$ fm).
The effective path length of a heavy quark in the QGP of life time $\tau_f$ is obtained as,
\begin{eqnarray}
L_{\mbox{eff}} &= \mbox{min}[\langle L \rangle, \frac{p_T}{m_T} \times \tau_f].
\end{eqnarray}
where  $m_T$ and $p_T$ are the transverse mass and transverse momentum of the heavy quark respectively.
We consider an isentropic cylindrical expansion as described in Ref.~\cite{EOS}. The temperature is estimated as a function of proper time using the  entropy conservation condition $s(T) V(\tau) = s(T_0 ) V (\tau_0)$ . 
The initial volume is calculated  by $V (\tau_0 ) = \pi [R_{tr}(N_{part})]^2 \tau_0$ and the transverse size $R_{tr}(N_{part})$ with number of participant $N_{part}$ is obtained as $R_{tr}(N_{part}) = R\sqrt{N_{part}/2A}$, with $A$ is mass number of the colliding nucleus.

 The initial and freeze-out times are taken as $\tau_0 = 0.3$ fm and $\tau_f = 6$ fm, respectively same as used in Ref.~\cite{Ours,KPV1}. Various other parameters used in our calculations are: initial temperature $T_{0} = 430$ MeV, $\langle L \rangle = 4.16$ fm, $N_{part} = 113$ and $\langle b \rangle = 9.68$ fm for minimum-bias $Pb-Pb$ collisions at centre of mass energy $\sqrt {s_{NN}} = 2.76$ TeV available at the LHC.

\subsection{Collisional energy loss: Brateen and Thoma (BT) formalism }
\label{Col}

The heavy quarks, while propagating through the QGP medium, collide elastically with the particles of the medium and lose energy. The collisional energy loss per unit length ($-dE/dx$) has been calculated in the past by several authors\cite{TG,BT,Alex,PP}. Brateen and Thoma\cite{BT} performed the most detailed calculation of $-dE/dx$ which was based on their previous QED calculation of $-dE/dx$ for muons\cite{muon}. The expression for $-dE/dx$ in the QGP medium of temperature $T$ for a relativistic heavy quark with mass $M_Q$ and energy $E\ll M_{Q}^{2}/T$ reads as\cite{BT},

\begin{eqnarray} 
-\frac{dE}{dx} & = &\nonumber \frac{8\pi\alpha_{s}^{2}T^{2}}{3}\left(1+\frac{n_f}{6}\right)\left(\frac{1}{v}-\frac{1-v^{2}}{2v^{2}}\ln{\frac{1+v}{1-v}}\right)\\
& &\times \ln\left(2^{n_f/(6+n_f)}B(v)\frac{ET}{m_{g}M_{Q}}\right) ,
\end{eqnarray}
where $\alpha_{s} = 0.3$ is the strong coupling constant, $B(v)$ is a smooth function of velocity ($v$) of heavy quark which can be taken approximately as 0.7, $m_{g} = \sqrt {(1+n_{f}/6)gT/3}$, is thermal gluon mass and $g = \sqrt {4\pi\alpha_{s}}$ and $n_f$ is the number of active quark flavors in the QGP medium. In the ultra-relativistic region i.e., $E\gg M_{Q}^{2}/T$, the expression for $-dE/dx$ becomes\cite{BT}

\begin{equation} 
-\frac{dE}{dx} = \frac{8\pi\alpha_{s}^{2}T^{2}}{3}\left(1+\frac{n_f}{6}\right) \ln\left(2^{n_f/2(6+n_f)}0.920\frac{\sqrt {ET}}{m_{g}}\right) 
\end{equation}

\subsection{Radiative energy loss: Reaction operator formalism (DGLV) }
\label{Rad}

Gluon radiation from a fast parton is the dominant and hence essential  mechanism of energy loss inside the QGP. The energy loss due to gluon radiation was first estimated in Ref.\cite{MG}. Latter many authors~\cite{mgm05,dokshit01,dead,wicks07,GLV,armesto2,B.Z,W.C,Vitev,AJMS,KPV} also calculated the radiative energy loss with many factors. The energy loss by gluon radiation from light quark jets in powers of gluon opacity ($L/\lambda$) (where $\lambda$ is the mean free path and $L$ is the path length traversed in the medium) has been calculated by using reaction operator formalism\cite{GLV}. This formalism was then extended to obtain the energy loss by gluon radiation from heavy quarks and simplified for the first order of opacity expansion~\cite{wicks07}. The expression for the average radiative energy loss from heavy quarks is given in the Appendix.

\subsection{Chromo-electromagnetic field fluctuations and energy gain}
\label{FF}

The collisional as well as radiative energy loss of heavy quarks inside QGP were obtained by considering the QGP medium in an average manner. However, QGP is a statistical system of mobile coloured charged particles (light quarks and gluons) which produce the chromo-electromagnetic fields due to their motion. Hence, QGP could be characterised by widespread stochastic chromo-electromagnetic field fluctuations. These field fluctuations are extremely important since they cause the fluctuation in the velocity of the propagating heavy quarks which are correlated with the fluctuations in the chromo-electromagnetic field. The presence of such correlations leads heavy quarks to gain energy.
 A quantitative estimation of the effect of these field
 fluctuations on the propagation of heavy quarks was done using the semiclassical approximation that is equivalent to the Hard Thermal Loop approximation based on weak coupling limit\cite{TG,BT}. 
 The estimated leading-log (LL) contribution 
 of the energy gain reads as\cite{Fl},
 \begin{eqnarray} 
 \left(\frac{dE}{dx}\right)_{\mbox{fl}}^{\mbox{LL}} = 2\pi C_F\alpha_{s}^{2}\left(1+\frac{n_f}{6}\right)\frac{T^3}{Ev^2}\ln{\frac{1+v}{1-v}} \ln{\frac{k_{\mbox{max}}}{k_{\mbox{min}}}},
 \end{eqnarray}
 where $k_{\mbox{min}} = \mu_g $ is Debye mass and
 $k_{\mbox{max}} = \mbox{min}\left[E, {2q(E+p)}/{\sqrt{M^2+2q(E+p)}} \right]$ with
 $q \sim T$ is the representative  momentum of the thermal partons (light quarks and gluons). The energy gain could be physically interpreted as the heavy quarks absorb gluons during their path of travel.

\begin{figure}[!h]
	\centering
	\begin{minipage}[b]{0.45\textwidth}
		\includegraphics[width=\textwidth]{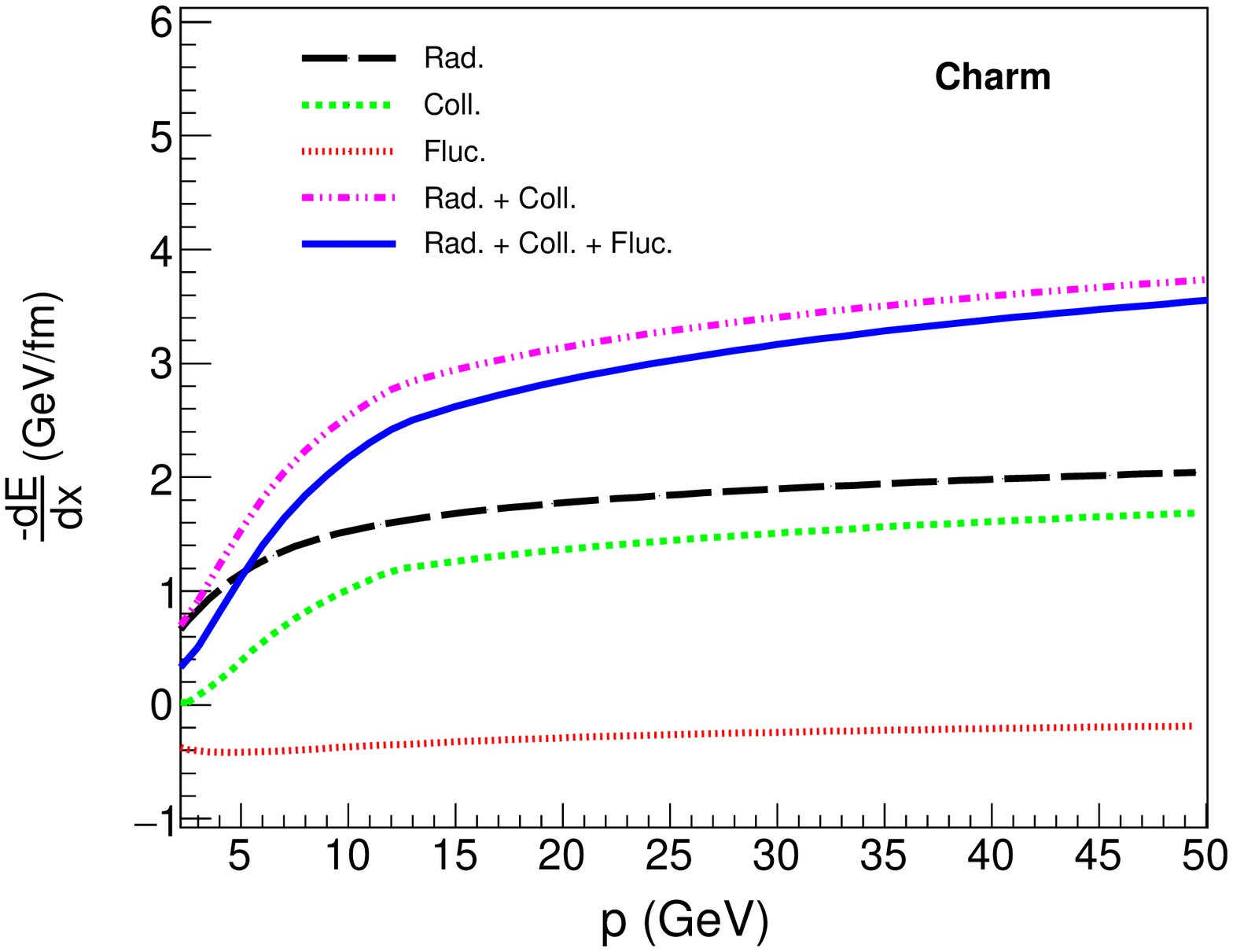}
		\caption{The energy loss of a charm quark inside the QGP medium as a function of its momentum, 
			obtained using BT\cite{BT}, DGLV\cite{wicks07,GLV} and Fluctuations\cite{Fl}.}
		\label{dedx_c}
	\end{minipage}
	\hfill
	\begin{minipage}[b]{0.45\textwidth}
		\includegraphics[width=\textwidth]{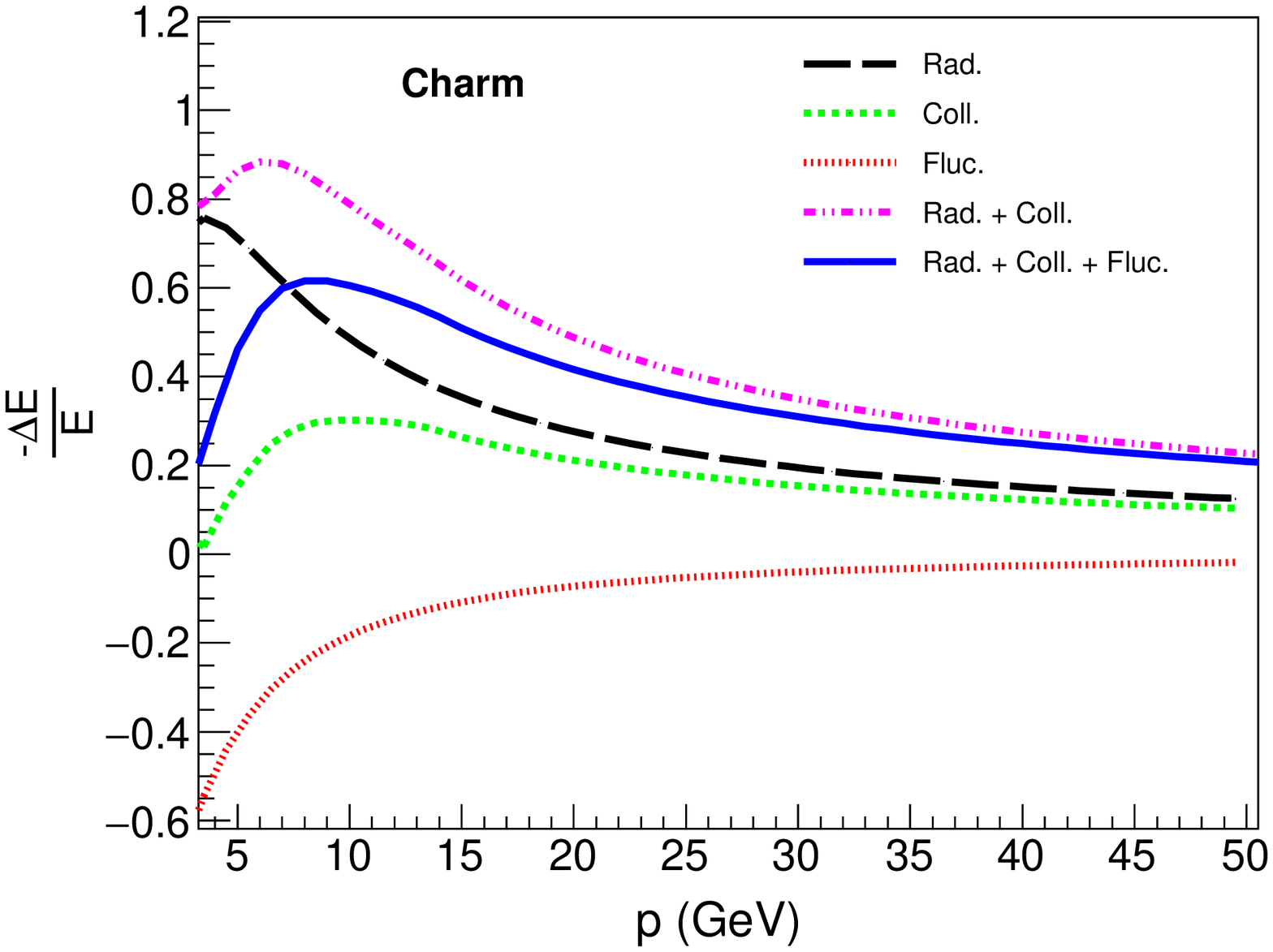}
		\caption{ Fractional energy loss of a charm quark inside the QGP due to BT\cite{BT}, DGLV\cite{wicks07,GLV} and Fluctuations\cite{Fl} as a function of its momentum. The path length considered here is $L=5$ fm.}
		\label{delEbyE_c}
	\end{minipage}
\end{figure}

The energy loss of heavy quarks as discussed in Sec.\ref{Col} and \ref{Rad} is calculated as a function of proper time. The initial conditions and medium evolution model used for the calculations are discussed in Sec.\ref{IC}. The calculated energy loss is then averaged over the temperature evolution of the QGP medium. The energy gain due to field fluctuations as described in Sec\ref{FF} has been also calculated in similar way.

We show the differential ($-dE/dx$) and fractional ($-\Delta E/E$) energy loss of a charm quark inside the QGP in figures \ref{dedx_c} and \ref{delEbyE_c} respectively. The effect of field fluctuations is also shown here. Our choice of  parameters are: number of flavours in the QGP medium, $n_f=2$; coupling constant, $\alpha_s = 0.3$ and charm quark mass, $M_c= 1.25$ GeV. It is observed that the energy loss increases with the momentum of charm quark however the rate of increment reduces at higher momenta. The DGLV radiative energy loss of charm quark is higher than the BT collisional energy loss. The energy loss is negative due to the field fluctuations which denotes energy gain. This energy gain is found to be important in the lower momentum region. It is due to the fact that the low momentum charm quarks are more affected by the field fluctuations and hence the energy gain of charm quarks becomes significant at the lower velocity limit. The effects of these fluctuations are significant to reduce the total energy loss. 

\section{Formalism of drag and diffusion coefficients}
\label{sec3}

The perturbative QCD (pQCD) calculations\cite{pQCD1,pQCD2} imply that the heavy quark thermalization time is larger than the light parton thermalization time scale, which  suggests that the heavy quarks are not in equilibrium  with the QGP medium, and hence the heavy quarks qualify to execute Brownian motion in the heat bath of light quarks and gluons. The Boltzmann transport equation is employed to describe such Brownian motion. Under the assumptions that the plasma is uniform, no external force present and the momentum change of heavy quarks due to collisions with the medium partons is relatively small, the Boltzmann transport equation reduces to Fokker-Plank (FP) equation. The FP equation reads as~\cite{BS,PR},

\begin{equation}
\frac{\partial f}{\partial t} = \frac{\partial }{\partial p_{i}} \left( p_{i}A(p)f + \frac{\partial }{\partial p_{i}} [B(p)f]\right),
\label{FPE}
\end{equation}

where $f$ is the phase space distribution function (here it is for heavy quarks), $p$ is the momentum of the heavy quarks. The quantities $A(p)$ and $B(p)$ are the usual drag and diffusion coefficients respectively.
Eq.\ref{FPE} can be used to study the evolution of heavy quarks in the QGP medium. During the propagation through the QGP, the heavy quarks lose energy via elastic collisions and bremsstrahlung gluon radiations. Along with that, heavy quarks gain energy due to the statistical field fluctuations of the QGP medium (see Ref.~\cite{Fl} for the details) which reduces the total energy loss of the heavy quarks. Therefore, the  estimation of drag ($A$) and diffusion ($B$) coefficients should include these energy losses as well as the energy gain processes. It should be mentioned here that the transport coefficients in FP equation (Eq.\ref{FPE}) usually evaluated for the collisional processes. However we consider the radiative processes also since they are the most dominating processes while a heavy quark moves very fast inside the QGP medium.

In this spirit, we use $-dE/dx$ to estimate the drag and diffusion coefficients of heavy quarks. $A_{Coll}$ and $B_{Coll}$ are drag and diffusion coefficient for collisional process. Similarly we define $A_{Rad}$ and $B_{Rad}$, ($A_{Fl}$ and $B_{Fl}$) for radiative, (field fluctuations) process. The effective drag and diffusion coefficients are defined as $A = A_{Coll} + A_{Rad} + A_{Fl}$ and $B = B_{Coll} + B_{Rad} + B_{Fl}$ respectively.  The net $-dE/dx$ is used to calculate the effective drag and diffusion coefficients. The effect of effective drag and diffusion on the heavy quark of momentum p in the QGP of temperature T can be defined as \cite{MGM,SKD10},

\begin{equation}
A =  \frac{1}{p} \left(- \frac{dE}{dx} \right)_{Coll+Rad+Fl}
\end{equation}
\begin{equation}
B = T \left(- \frac{dE}{dx} \right)_{Coll+Rad+Fl}
\end{equation}
respectively. These effective transport coefficients are important quantities containing the dynamics of energy loss and gain processes of the heavy quarks in the QGP medium. One can average out $A$ and $B$ over momentum, implying that the dynamics is dominated by the energy loss and gain processes in the heat bath.

 We use power-law distribution and differential energy loss calculations BT~\cite{BT} and DGLV~\cite{wicks07,GLV} to perform the momentum averaging of $A$ and $B$. To sample the initial transverse momentum of charm quarks, we use the following power-law parametrization~\cite{plbass}:
 
  \begin{equation}
  \frac{dN}{d^{2}p_{T}}  \propto  \frac{1}{(p_{T}^{2}+\Lambda^{2})^{n}} 
  \end{equation}
  where, $\Lambda = 2.1$ and $n = 3.9$.
  The time dependence in $A$ and $B$ comes from assuming the temperature $T$ is decreasing with time as the system expands cylindrically in isentropic nature. For time averaging of $A$ and $B$, we calculate $A$ and $B$ from energy loss for different times during the expansion of the system and then averaging them over the entire evolution of the system.

\begin{figure}[!h]
	\centering
	\begin{minipage}[b]{0.45\textwidth}
		\includegraphics[width=1.1\linewidth]{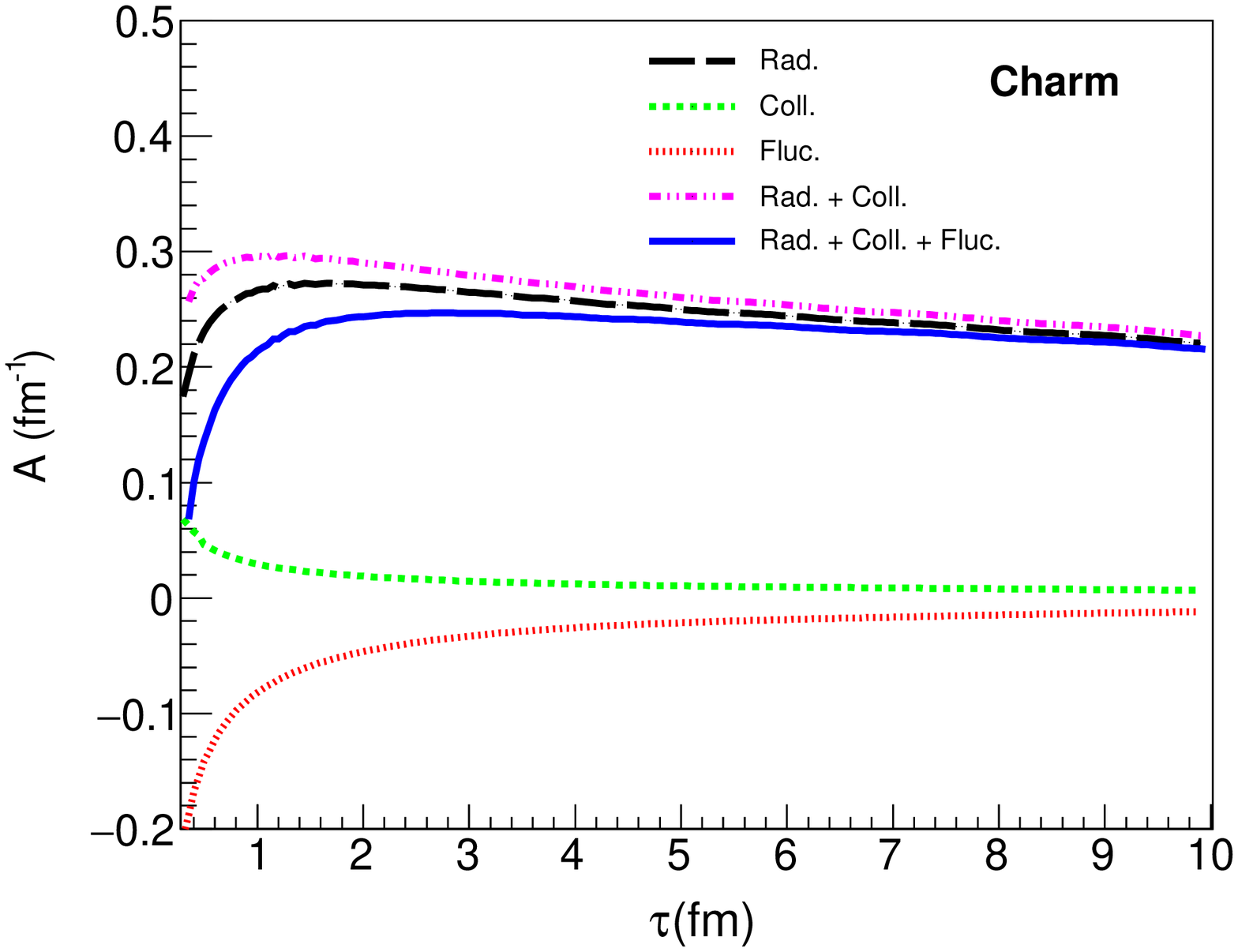}
		\caption{The drag coefficient of a charm quark inside the QGP medium as a function of time, obtained for different energy loss schemes (BT~\cite{BT} and DGLV~\cite{wicks07,GLV}) along with the effect of fluctuations~\cite{Fl}. }
		\label{drag_tau}
	\end{minipage}
	\hfill
	\begin{minipage}[b]{0.45\textwidth}
		\includegraphics[width=1.026\linewidth]{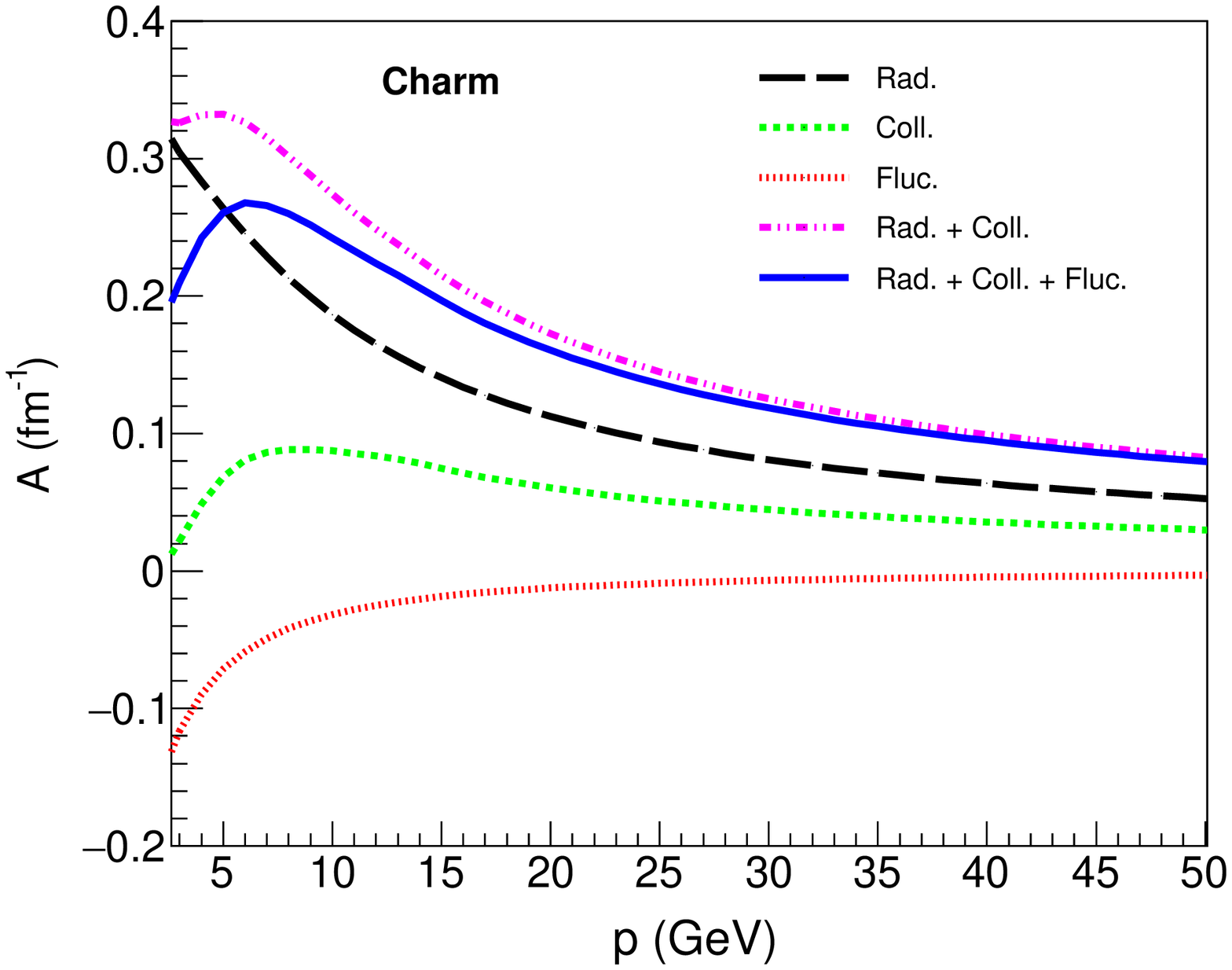}
		\caption{The drag coefficient of a charm quark inside the QGP medium as a function of its momentum, 
			obtained for different energy loss schemes(BT\cite{BT} and DGLV\cite{wicks07,GLV}) along with the effect of fluctuations\cite{Fl}.}
		\label{drag_p}
	\end{minipage}
\end{figure}

In figures  \ref{drag_tau} and \ref{drag_p}, the variation of drag coefficients ($A$) of a charm quark with time and charm quarks momentum have been depicted respectively. The values of $A$ are positive and decrease with time and momentum where only the energy loss processes have been considered. For the field fluctuations, as these fluctuations cause charm quarks to gain energy~\cite{Fl,Ours}, the values of A are negative. We also observe that the contribution of radiative energy loss is large compared to the collisional one which is consistent with the findings of Das et. al.~\cite{SKD10}. The total contribution of radiative and collisional losses ($Coll. + Rad. $) are large enough whereas the inclusion of the effect of field fluctuations decreases the total drag coefficient ($Coll. + Rad. $) and we call it effective drag coefficient ($Coll. + Rad. + Fluc.$). The reduction of drag coefficient is more at the lower momentum region ($p < 25$ GeV) since the fluctuations are significant at lower momentum as discussed earlier. At $p > 25$ GeV region, the effect of the fluctuations on drag coefficient is negligibly small.

\begin{figure}[!h]
	\centering
	\begin{minipage}[b]{0.45\textwidth}
		\includegraphics[width=1.1\linewidth]{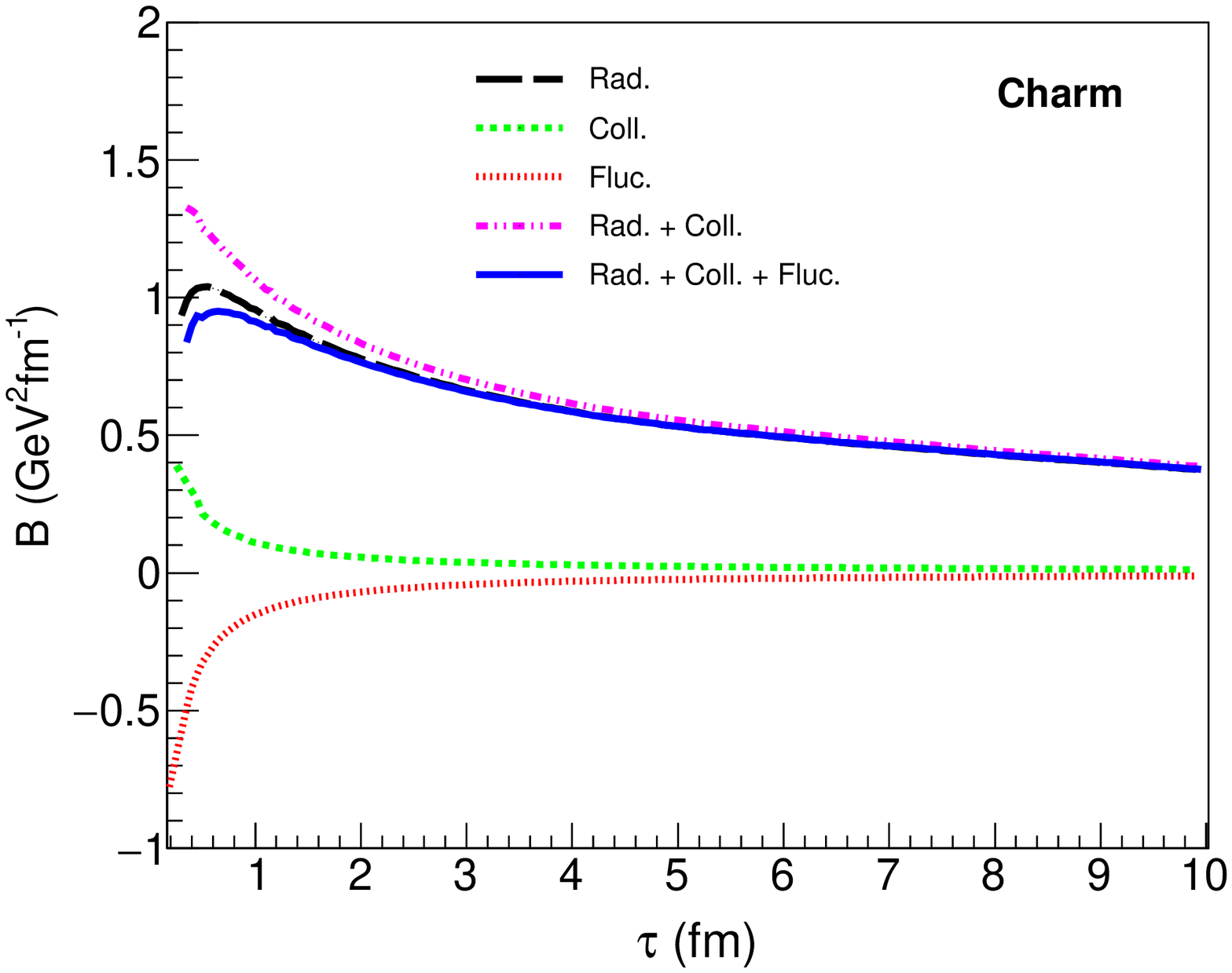}
		\caption{The diffusion coefficient of a charm quark inside the QGP medium as a function of time, 
			obtained for different energy loss schemes(BT~\cite{BT} and DGLV~\cite{wicks07,GLV}) and along with the effect of fluctuations~\cite{Fl}. }
		\label{diff_tau}
	\end{minipage}
	\hfill
	\begin{minipage}[b]{0.45\textwidth}
		\includegraphics[width=1.026\linewidth]{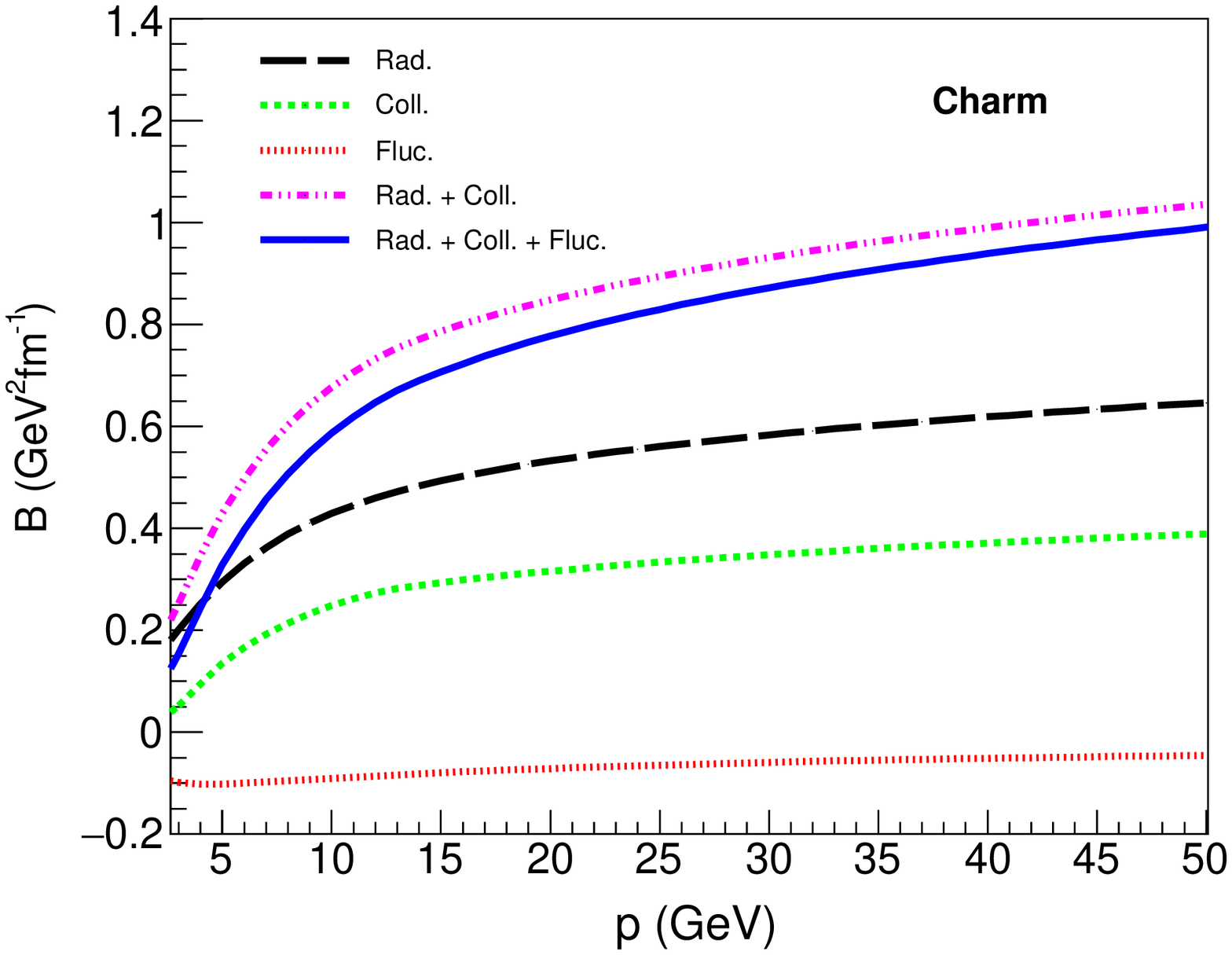}
		\caption{The diffusion coefficient of a charm quark inside the QGP medium as a function of its momentum, 
			obtained for different energy loss schemes(BT~\cite{BT} and DGLV~\cite{wicks07,GLV}) and along with the effect of fluctuations~\cite{Fl}.}
		\label{diff_p}
	\end{minipage}
\end{figure}

Once the drag coefficient is averaged out over momentum using the properties of heat bath (as discussed earlier), the diffusion coefficient can also be averaged out over momentum since it is derivable from the drag coefficient through Einstein's relation. Figures \ref{diff_tau} and \ref{diff_p} display the diffusion coefficients ($B$) of a charm quark as a function of time and charm quarks momentum respectively. We observe that the values of $B$ are positive and decrease with time  and increase with momentum when only the energy loss processes are considered. The values of $B$ are negative for field fluctuations because the charm quarks gain energy due to field fluctuations. It results a reduction of total diffusion coefficient ($Coll. + Rad. $) to an effective diffusion coefficient, $B_{eff}$  ($Coll. + Rad. +Fluc.$) which is less compared to $Coll. + Rad.$ case.

\section{Shear viscosity to entropy density ratio ($\eta/s$) of the QGP probed by charm quarks}
\label{sec4}

Viscosity measures the resistance of a fluid deformed either by tensile stress or shear stress. The less viscosity causes  greater fluidity. In order to characterize QGP, amongst many,  $\eta/s$ is one of the important quantity. It is an important dimensionless measure of how imperfect or dissipative the QGP is. 

A  heavy quark with certain momentum while propagating through the QGP medium encounters the medium partons and hence the momentum exchange occurs with the medium partons. The momentum exchange results minimization of momentum gradient in the system. Hence, it is related to the shear viscous coefficients of the system which drives the system towards a reduced momentum gradient. The expression of $\eta/s$ has been calculated in Ref.~\cite{Abhijit}, which reads as: 

\begin{equation}
\frac{\eta}{s} \approx  1.25 \frac{T^{3}}{\hat{q}} ,
\label{etaq}
\end{equation}
where $T$ is the temperature of the medium and $\hat{q}$ is the transport coefficient which is defined as square of the average exchanged momentum between the heavy quark and bath particles per unit length. During the interactions of heavy quark with bath particles, the momentum diffusion occurs in the medium which is expressed through the diffusion coefficient. The diffusion coefficient and $\hat{q}$, both are directly related to the momentum transfer and hence $\hat{q}$ is proportional to diffusion coefficient which is more often used in the diffusion equation as discussed in Ref.\cite{Abhijit,Xu}. The $\hat{q}$ in Eq.\ref{etaq} is defined for gluon~\cite{Abhijit}. We have considered the color factor difference while relating the charm quark diffusion coefficient $B$ with $\hat{q}$. The relation between $B$ and $\hat{q}$ is $B=\hat{q}/4$.
Eq.\ref{etaq} has been used to estimate $\eta/s$ of the QGP medium in the light of heavy quarks energy loss by Mazumder et. al.\cite{Sur}. In this work, we estimate $\eta/s$ from Eq.\ref{etaq}, where $\hat{q}$ is obtained from the effective diffusion coefficient which takes into account the effect of the field fluctuations.

\begin{figure}[htb!]
	\centering
	\includegraphics[scale=0.4]{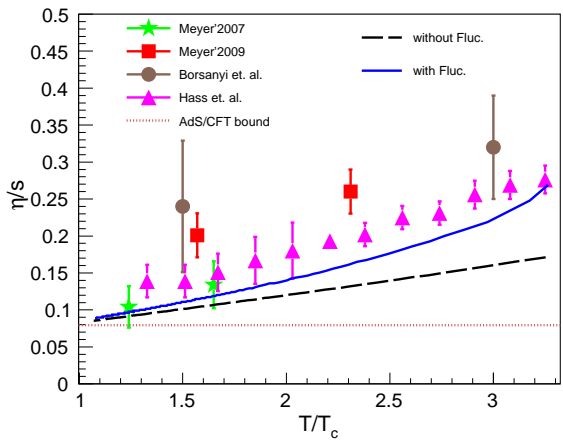}
	\caption{The viscosity to entropy density ratio ($\eta/s$) as a function of $T/T_c$, is compared with the results obtained by LQCD calculations\cite{Mayer07,Mayer09,BorsanyiLQCD14} and functional renormalization group calculations\cite{Haas}. Here critical temperature $T_c$ is taken 155 MeV.}
	\label{etabys}
\end{figure}

In Figure \ref{etabys}, we display $\eta/s$ as a function of $T/T_c$ when the charm quarks undergo both collisional and radiative processes along with the chromo-eletromagnetic field fluctuations which cause the charm quarks to  gain energy. We observe that the effect of such fluctuations increases the values of $\eta/s$ of the QGP. The obtained values of $\eta/s$ are close to the LQCD calculations\cite{Mayer07,Mayer09,BorsanyiLQCD14} and findings of functional renormalization group technique\cite{Haas} within their uncertainties when we include the effect of field fluctuations into account. 

AdS/CFT calculations give a lower bound of $\eta/s$, which is $\frac{\eta}{s} \geq \frac{1}{4\pi}$. The obtained $\eta/s$ values go slightly below this AdS/CFT lower bound near $T_{c}$ which might be unphysical. It is hard to characterize the QGP near critical point. However it is worth mentioning that the theoretical uncertainties may appear in our calculations due to thermalization of the medium which may occur due to uncertainties in initial conditions. Mean energy loss calculations used here is based on semi-classical approximation which is equivalent to the Hard Thermal Loop approximation on the basis of weak coupling limit. The non-Abelian terms in the QCD equations of motion is also ignored here. Thus the uncertainty in the estimation of $\eta/s$ from the energy loss calculation may arise.

\section{Summary and Conclusion}
\label{sec5}
The  energy loss suffered by an energetic heavy quark inside QGP medium provides the dynamical properties of the QGP which are reflected in the nuclear modification factor of heavy-flavoured mesons. In the phenomenological study of heavy flavour suppression, the effect of field fluctuations along with the energy loss processes is important to describe the measured suppressions. In this article the effect of such fluctuations on charm quarks drag, diffusion coefficients and $\eta/s$ of the QGP medium have been investigated. It is observed that the effect of the fluctuations reduces the drag and diffusion coefficients compared to the total contributions from collisions and gluon radiations. The radiative loss is dominant over the collisional counter part in drag and diffusion coefficients. The effect of the fluctuations has significant impact on $\eta/s$ of the QGP medium. These fluctuations increase the values of $\eta/s$. The obtained values of $\eta/s$ close to the LQCD and functional renormalization group calculations when the effect of the fluctuations are taken into account.

\section*{Acknowledgements}
We acknowledge Prof. J. Alam and Prof. Munshi G. Mustafa for fruitful discussions.

\section*{Appendix: Reaction operator formalism (DGLV)}
\label{App}

The average energy loss due to gluon radiation from a heavy quark, obtained by the reaction operator formalism is written as\cite{KPV},

\begin{equation}
\frac{\Delta E}{L} = E\frac{C_F\alpha_{s}}{\pi}\frac{1}{\lambda}\int^{1-\frac{M_Q}{E+p}}_\frac{m_g}{E+p} dx \int^\infty_0 \frac{4\mu_{g}^{2}q^{3}(X \ln Y + Z)}{(\frac{4Ex}{L})^2+(q^2+\beta^2)^2} dq ,
\end{equation}

where

\begin{equation}
\beta^2 = m_{gt}^2(1-x)+M_Q^2x^2, \lambda^{-1} = \rho_g\sigma_{Qg} + \rho_q\sigma_{Qq}
\end{equation}
with $\rho_g = 16T^3 \frac{1.202}{\pi^2}$ and $\rho_q = 9n_fT^3 \frac{1.202}{\pi^2}$ are the densities of quarks and gluons in the QGP medium respectively where $\sigma_{Qq} = \frac{9\pi\alpha_s^2}{2\mu_g^2}$ and $\sigma_{Qg} = \frac{4}{9}\sigma_{Qq}$. Here $C_F = 4/3$ which determines the coupling strength between the heavy quark and gluon and $m_{gt} = \mu_g/\sqrt 2$ is the transverse gluon mass.

The functions $X$, $Y$ and $Z$ are given by the following relations:
\begin{equation}
X = \frac{2\beta^2}{f_{\beta}^3} (\beta^2+q^2)
\end{equation}

\begin{equation}
Y = \frac{(\beta^2+K)(\beta^2Q_{\mu}^{-}+Q_{\mu}^{+}Q_{\mu}^{+}+Q_{\mu}^{+}f_{\beta})}{\beta^2\big(\beta^2(Q_{\mu}^{-} - K)-Q_{\mu}^{-}K+Q_{\mu}^{+}Q_{\mu}^{+}+f_{\beta}f_{\mu}\big)}
\end{equation}

\begin{eqnarray}
Z = \frac{1}{2q^2f_{\beta}^2f_{\mu}}[\beta^2\mu_g^2(2q^2-\mu_g^2)+ \beta^2(\beta^2-\mu_g^2)K+  \nonumber \\ Q_{\mu}^{+}(\beta^4-2q^{2}Q_{\mu}^{+})+f_{\mu}(\beta^2(\mu_g^{2}-\beta^2-3q^2)+2q^{2}Q_{\mu}^{+})+3\beta^{2}q^{2}Q_{k}^{-} ]
\end{eqnarray}

Where, $K = k_{max}^2 = 2px(1-x)$, $Q_{\mu}^{\pm} = q^2 \pm \mu_g^2$, $Q_{k}^{\pm} = q^2 \pm k_{max}^2$, $f_{\beta} = f(\beta,Q_{\mu}^{+},Q_{\mu}^{-})$ and $f_{\mu} = f(\mu_g,Q_{k}^{+},Q_{k}^{-})$ with $f(x,y,z) = \sqrt{x^4+2x^2y+z^2}$.

\end{document}